\documentstyle[preprint,aps,prl,epsbox]{revtex}
\tightenlines
\begin{document}
\draft
\title{Low Energy Solar Neutrino Detection by using Liquid Xenon}
\date{\today}
\maketitle

\begin{center}
\newcounter{foots}
\vspace{1cm}

Y.Suzuki\\

(for the Xenon Collaboration~\cite{Xenon-Coll})\\

%
\footnotesize \it
Kamioka Observatory, Institute for Cosmic Ray Research, 
University of Tokyo, Higashi-Mozumi, Kamioka,
Gifu 506-1205, Japan\\
\normalsize
%
%
%
%
%
%
%
\vspace{2cm}
(Talk presented at LowNu workshop, June-15-2000, Sudbury, Canada)\\
\end{center}

\begin{abstract}

Possibility to use ultra pure liquid Xenon as a low energy solar neutrino 
detector by means of $\nu$+e scatterings is evaluated. 
A possible detector with 10 tons of fiducial volume will give 
$\sim$14 events for pp-neutrinos and $\sim$6 events for $^{7}$Be neutrinos 
with the energy threshold at 50 keV. 
The detector can be built with known and established technologies.
High density of the liquid- Xe would provide self-shields against
the incoming backgrounds originating from the container and outer environments.
Internal backgrounds can be reduced by distillation and other techniques. 
Purification of the liquid Xe can be done continuously throughout 
the experiment. 
The spallation backgrounds are estimated to be small though an experimental
determination is neccessary.
The liquid-Xe detector can also provide 
a significantly better sensitivity for the double beta decay and a dark matter 
search. However the 2$\nu$ double beta decay of $^{136}$Xe would be most
background. 
It could be overcome if the 2$\nu$ lifetime is longer than 10$^{22}$yr.
However, an isotope separation
of $^{136}$Xe is inevitable for a shorter lifetime.

The isotope separations would, intoroduce a new opportunity to definitively 
identify dark matter.  The interesting feature 
in addition to the solar neutrino measurements will also be discussed.

\end{abstract}

\newpage

\narrowtext
\section{Introduction}


The long standing solar neutrino problem is now widely thought to be caused 
by neutrino oscillations, since the discovery of neutrino 
oscillation in the study of the atmospheric neutrinos in 1998 
clearly demonstrated that neutrinos have masses~\cite{SK-Atm-Discovery}.
The on-going real-time solar neutrino experiments, 
Super-Kamiokande~\cite{SK-Solnu}, SNO~\cite{SNO-Solnu}
would provide an unique oscillation parameters in near future.
However, the $^{8}$B neutrinos where both Super-Kamiokande and SNO detector can detect,
are the very fractional tail (0.17\%) of the entire solar neutrino spectrum.
It is very important and necessary to establish the neutrino oscillation by 
the low energy solar neutrinos, namely by pp and $^{7}$Be neutrinos.

No experiments have separately measured the pp and $^{7}$Be neutrino flux 
so far. 
The gallium(SAGE, GALLEX, GNO)~\cite{Ga-exp}. and chlorine(Homestake)~\cite{Cl}experiments have measured the integrated flux above a certain energy, 235keV 
for Ga and 817keV for Cl.
Borexino~\cite{Borexino} in very near future would determine a $^{7}$Be-flux 
separately by means of neutrino electron scattering.
The neutrino electron scattering (neutral current and charged current) and the
charged current (absorption) measurements provide complimentary 
information: the
information on charged and neutral current interaction rates can be obtained 
separately from the two measurements, which is necessary for the compete
understanding of neutrino oscillation phenomena including a possible case for
the oscillation to sterile neutrinos.
LENS~\cite{LENS} that utilizes the neutrino absorption by Yb, aims 
to provide a charged current information of the $^{7}$Be and pp neutrinos.

A few ambitious proposals, HERON~\cite{HERON}, HELLAZ~\cite{HELLAZ}, 
NEON~\cite{HeNe} and so on, 
are under consideration to detect pp-neutrinos by means of the neutrino 
electron scattering.
 
We consider here a liquid-Xe detector to measure both pp and $^{7}$Be
neutrino spectrum in real time utilizing the neutrino electron elastic 
scattering of which we know the cross section very well.
The characteristics of the liquid-Xenon is very well known and it works as a 
very good scintillation material.
 
The $\nu$+e scattering experiments give relatively higher statistics
than the absorption experiments. But no coincidence information
is available, and therefore they require ultra-pure materials and environments.

The ultra-low backgrounds have been studied extensively so far, 
especially by those working for solar neutrino experiments.
Borexino group~\cite{LowBG-I} has demonstrated at their CTF 
facility that the contamination level of a few times 10$^{-16}$g/g for 
U/Th and 10$^{-18}$ for the ratio $^{14}$C/$^{12}$C has possibly been 
achieved in
their liquid scintillator.
This is the level required for the $^{7}$Be- neutrino measurement by the
$\nu$+e scattering experiment.
Borexino is, however limited by the contamination of $^{14}$C to go down to 
the pp-neutrino region. This is an intrinsic limit of the detector which uses
organic scintillators.

The Xe detectors do not contain carbon and therefore have
less problem due to $^{14}$C. 
Therefore there is a possibility to 
go down to the pp-neutrino regions with a similar background
level required for Borexino.

\section{Detection of Solar neutrinos by means of Liquid-Xenon}

\subsection{Liquid-Xenon Scintillators}

Liquid Xenon is a good scintillator. The scintillation mechanism is 
well known. Xe atoms are either exited or ionized
when charged particles have passed through Xe.
Both the excitation and the 
ionization following the recombination produce the excited Xe$^{*}_{2}$
states which in turn give ultra-violet photons at 174nm. 
The de-excitation time
is 3ns for the singlet states and 27 ns for the triplet states.
The recombination of the ion and electron will take place with a characteristic
time determined by the ionization density. If excited by electrons, the 
recombination time is typically 40nsec and if exited by alphas, then the 
recombination time is about 3 nsec. This may allow separation of electron
from nuclei by the pulse shape discrimination.

The Xe scintillation produces about 43,000 photons per MeV that is about 70\% 
of NaI.
The light attenuation length is not well known and depends upon the level of
purity. 
It is supposed to be longer than 1m for high purity liquid 
Xenon~\cite{muegamma}. 
The Rayleigh scattering length was measured by a few groups and found to
be about 30cm. But this also depends on the purity of the liquid.

The ultra-violet photons can be detected directly by photomultiplire tubes
(PMTs).
No wave length shifter is needed while a 
wave length shifter is needed for the liquid-He and liquid-Ne 
detectors~\cite{HeNe} where $\sim$80nm photons are emitted.

The PMTs can be placed either
in the liquid or outside of the detector through the windows.
One of the photo-tube (for example, R6041Q/HAMAMATSU) can be operated at 
about -108 $^{o}$C, and 
has a quartz window of effective diameter of 46mm$\phi$. 
The wave length of 174nm is just around the edge of cut off wave length of 
the quarts window. 
The more the thickness increases, the more you lose the light.
The quantum efficiency is 10$\sim$15\%. 

The PMTs that are sensitive to ultra-violet photons operated at room temperature
are commonly obtained.
One of them (for example, EMI9426B) has very high QE of 35\% for 174nm photons with 
MgF$_{2}$ 2 inch-window. 
The MgF$_{2}$ window with 12mm thickness has a more than 95\% transmission for 
174nm photons.

Comparisons among noble gas scintillators are listed in 
table~\ref{tbl:comp}.

The operating temperature of 165K (-108$^{o}$C) for liquid Xe is highest 
among them, which makes the cryogenic very simple and easy.
A single layer of vacuum shield is sufficient to keep the detector operated.
It requires very low power to keep Xenon in the liquid phase.
The 800 litter(2.4ton) detector proposed for the $\mu$$\rightarrow$$e$+$\gamma$
experiment\cite{muegamma} can be kept only by 200W.
One of the advantage is that Liquid-Nitrogen can be used to liquefy Xe and
this makes the transfer of the Xe through the cryogenic system very easy.

The high density ($\rho$=3.06g/cm$^{3}$) and high atomic number (A=131) would
provide a good self-shielding. 
The radiation length and nuclear collision length is about 2.7cm and 33.6cm,
respectively. For example additional thickness of 30cm corresponds to
the 4m of the water shields for the electro-magnetic components of backgrounds.
Therefore the requirements for the purity of the structure and environmental 
backgrounds are less serious than those for the other noble gas scintillators.

Electrons and ions are known to drift in the liquid Xenon.
The drift velocity of electrons is $\sim$2$\times$10$^{5}$cm/s when 200/V/cm 
of the electric field is applied. In reference~\cite{Electron-Drift}, an electron drift up to 2m 
has been observed. 
Positive ions can also drift. The mobility of the ions are measured to 
be $3\times$10$^{-4}$cm/V/s.  This brings the idea of the 'Ion Sweeper' that 
will be discussed later.

Xenon has a very unique feature among other noble gas scintillators. 
It produces proportional scintillation throught a multiplication process in an
electric field~\cite{Xe-Prop-Scin}.
In the liquid, the field strength above 5$\times$10$^{5}$V/cm is required in
order to produce a 
proportional light. Very thin wires
with a diameter of less than 10 $\mu$m is necessary for this condition.
A few photons per electron are expected to be emitted reduced.
The proportional scintillation is much prominent in gas phase even with 
an electric
field about two orders of magnitude smaller than that in the liquid.
The wave length of the proportional light is also 175nm.

The particle discrimination between electrons and alphas (or nuclei) is
possible by the pulse shape discrimination~\cite{PSD-I} that is related to the
emission mechanism mentioned above. The characteristic time due to the 
different recombination times is 3 nsec for nuclei and 40 ns for 
electrons.

The proportional scintillation light would provide another way to discriminate
electrons from nuclei. 
If an electric field applied, those electrons produced through the ionization 
by electrons/$\gamma$'s, not by nuclei, start to drift before the 
recombination takes place. Those drifted electrons can be guided into a region
with a high electric field, then the proportional 
scintillation lights can be emitted. Thus the discrimination between electrons
and other particles becomes possible. In the liquid, however, the amount of 
light emitted is a few photons per electron.

If those electrons are guided into the gas phase, then the more intense
proportional light can be emitted even in a lower electric 
field~\cite{Liq-Gas}.
The potential to pull electrons into gas phase ($\sim$0.7eV) can be obtained
by applying the escape field of 2$\sim$3kV/cm. 
Consequently, one gets large proportional lights in the gas phase and small 
signal in the liquid Xe for electrons and one gets a signal only in the liquid
for alpha or other nuclei.

This method could identify the $\alpha$ background originating from the U/Th 
decay chain and from possible dark matter interactions in the detector.
 
The prototype of such a  detector---called two phase detector---has been
developed for a dark matter search~\cite{SSuzuki-I}.

The liquid Xe has been studied extensively for many years and been widely used
for dark matter searches and for studies of double beta decay.
Therefore it is a sort of an established and known technology.

In order to detect about 10$\sim$20 pp-$^{7}Be$ neutrinos per day, 10 tons 
(fiducial volume) of liquid Xenon are needed. This amount is not a big 
surprise. There is an approved experimental project to use 
800 litter(2.4tons) of liquid Xenon for
$\mu\rightarrow e+\gamma$~\cite{muegamma} detection.

A most problem we have to carefully address is backgrounds, which will be 
discussed in later sections.

\subsection{Solar Neutrino signal}

\begin{figure}
\begin{center}
\psbox[height=9.5cm]{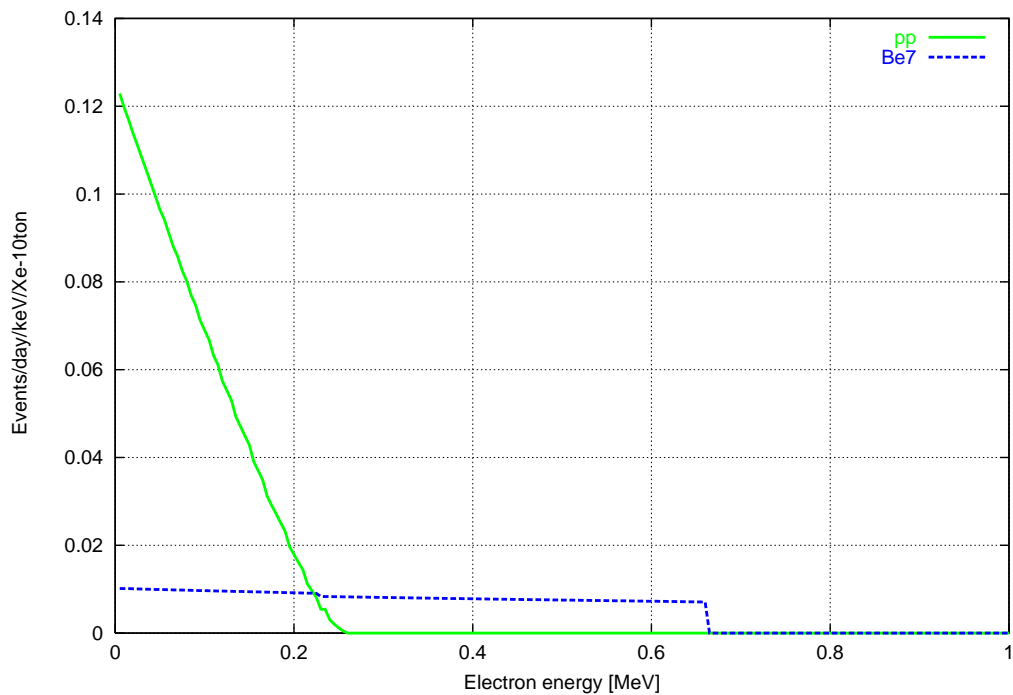}
\caption{The recoil electron energy spectrum.}
\label{fig:Spec}
\end{center}
\end{figure}

\begin{figure}
\begin{center}
\psbox[height=9.5cm]{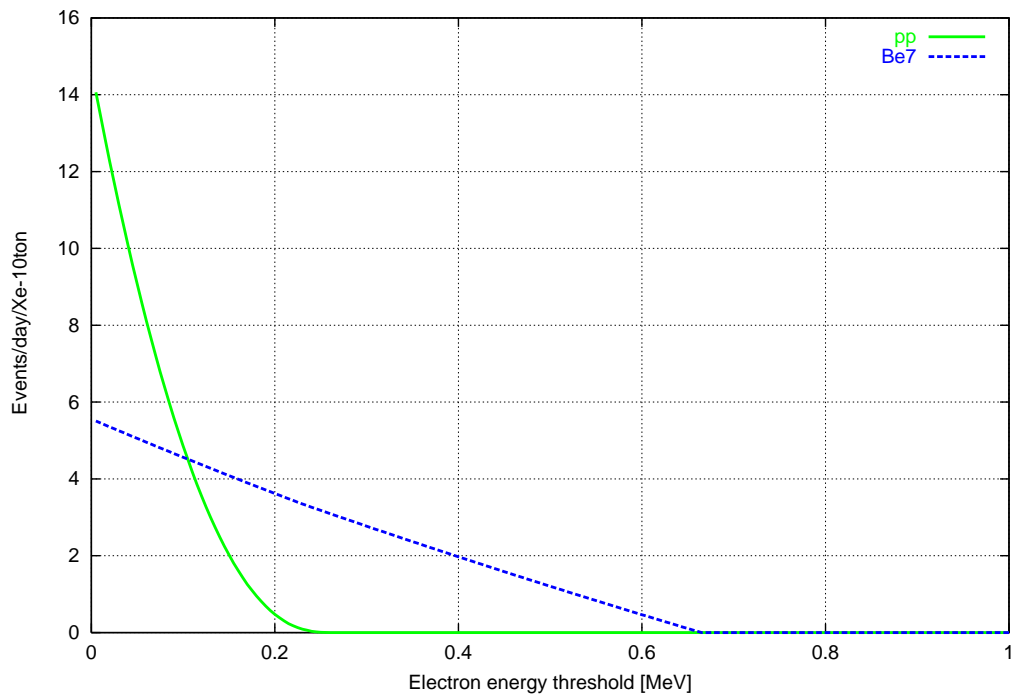}
\caption{The expected event rate as a function of the threshold energy.}
\label{fig:XeRate}
\end{center}
\end{figure}

The cross section of the $\nu+e\rightarrow\nu+e$ scatterings is 
very well known. Therefore we do not need a calibration for the absorption
(interaction) cross sections.
The flux averaged cross sections for pp- and $^{7}$Be neutrinos are 
1.2$\times$10$^{-45}$cm$^{2}$ and 61$\times$10$^{-45}$cm$^{2}$, respectively.
The Compton like maximum energy of pp-neutrinos is 260keV.
The energy spectrum expected for no oscillations is shown in 
Fig.~\ref{fig:Spec} without energy resolution smearing.
The event rate expected as a function of the threshold energy is shown 
in Fig~\ref{fig:XeRate}.
We expect $\sim$14 pp-neutrino events/day and $\sim$6 $^{7}$Be neutrinos/day for 10 tons 
of liquid Xe. 

Liquid Xenon are considered also as a solar neutrino detector by means of the 
neutrino absorption process of $^{131}$Xe with the threshold energy of 352keV 
that is mostly sensitive to the $^{7}$Be neutrinos~\cite{Xe-inverse}. 
However the event rate is about 1500 events/yr/kton.
For 10 ton detector, the events from this reaction is negligible.

\subsection{detector}

\begin{figure}
\begin{center}
\psbox[height=12.0cm]{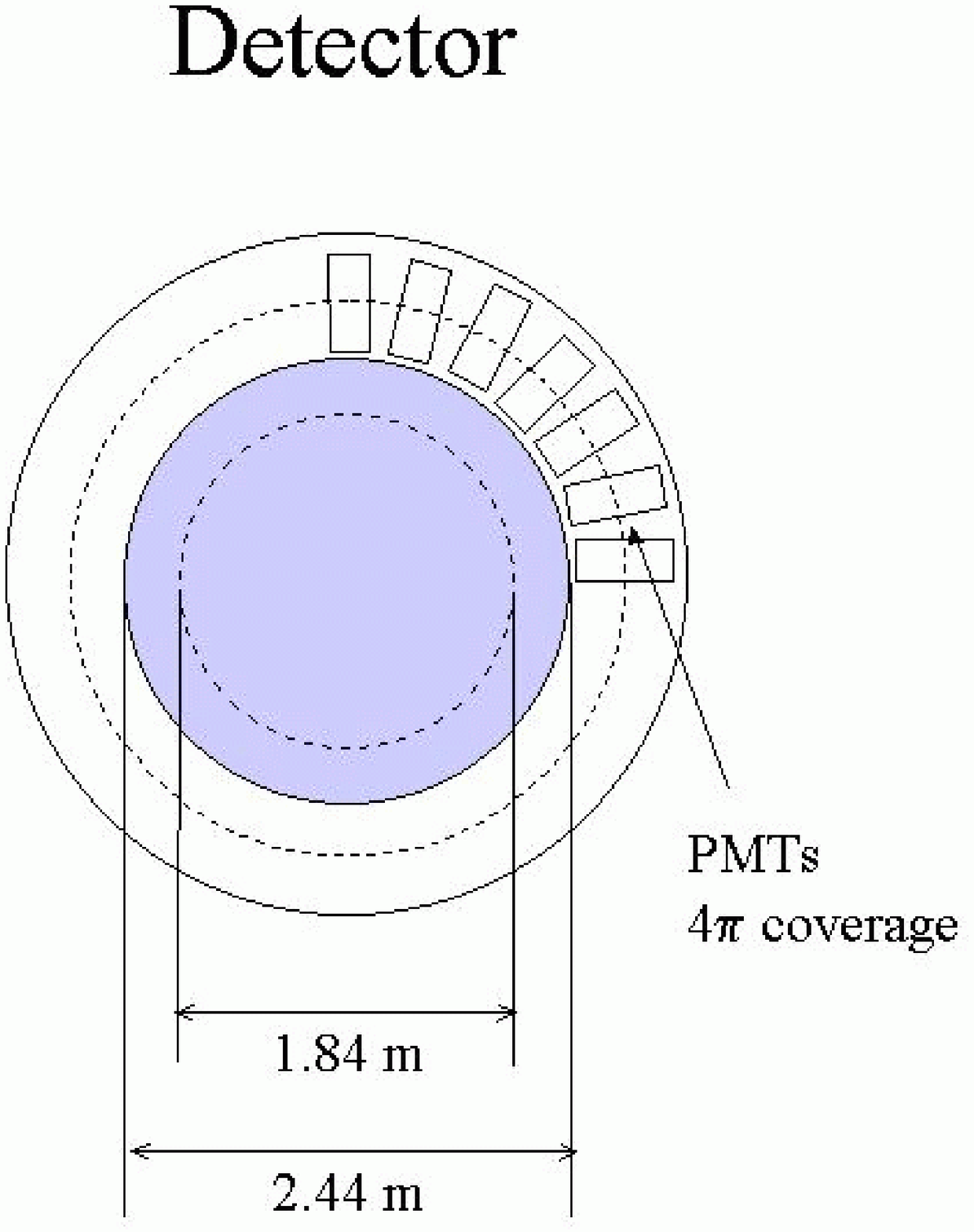}
\caption{Single phase detector.}
\label{fig:sp-detector}
\end{center}
\end{figure}

Two types of the detector can be considered---single phase (liquid) Xenon
detector and double phase (liquid+gas) detector.
\paragraph{single phase detector}
A schematic view of the single phase detector is shown in 
Fig.~\ref{fig:sp-detector}.
It is a spherical counter and contains 23.27 tons of liquid Xe in the 
inner container with a radius of 1.22m. 
The radius for the 10 tons fiducial volume is 0.92m and an additional 30cm 
layer act as self-shielding against the backgrounds entering into the inner 
container.
This provide the shield of 11 radiation length, which is equivalent to 
the 4m of the water shield for $\gamma$'s and electrons.

Phototubes can be placed either in the liquid or out side of the liquid through
quartz or MgF$_{2}$ windows. 
If you place the PMTs 
in the liquid Xe, you need additional insensitive layer filled with liquid Xe. 
For this schematic design, we have placed all the photo-tube outside of the 
inner container.
The inner container can be made either by stainless steel, oxygen free copper,
or other materials, depending on the contamination level of U/Th.
The windows may be made by MgF$_{2}$, quartz or other suitable materials.

For the heat insulation purpose, there is a gap between windows and the PMT 
surface.
The layers having an array of PMTs act also as a vacuum insulator.

Suppose we cover 40\% of the surface with PMTs at the distance
of 1.22m from the center, we need either 1640 inch diameter PMTs or 3690
PMTs with 2 inch diameter.

Outside of the PMT layers, the water layer of 2 m thickness to provide the 
shields against backgrounds from environments is placed. If necessary, the 
thickness of this water shield can be increased. This layer can be made as an 
active veto like Super-Kamiokande anti-counters. 

For the small scale test experiment~\cite{muegamma} with 2.3$\ell$ of active 
volume of which 
$\sim$40\% of the surface are coverd by PMTs, the energy resolutions of 9.4\% 
and 7\%, for 320 and 662 keV $\gamma$- rays were obtained~\cite{muegamma}. 
The extrapolated
resolute at 100keV is 17.5\%. They claim that they could improve the collection
factor about 2 in future. 
The event vertex can be reconstructed by the pulse height distributions and the
timing distributions. For the small size test experiment, 
they obtained 7mm for 320 keV~\cite{muegamma}.
Pulse shape discrimination of electrons and alphas (nuclei) may be possible in the
single phase detector.
However, detailed study for this possibility is necessary.

\paragraph{two phase detector}
Another interesting configuration is that we use both liquid and gas phases. 
This kind of arrangment has been proposed and studied 
extensively~\cite{Two-Phase,SSuzuki-I} 
for the dark matter experiments in order to
obtain better discrimination of electrons from nuclear recoils.
A few hundred V/cm of the electric field makes electrons produced by the 
ionization to drift towards the anode at the surface of the liquid, while the 
electrons produced by the 
ionization of alpha particles to recombine.
Therefore we expect larger prompt scintillation lights for the alpha particles
and a small signal for the electron events in the liquid. 

If the field of about 2 to 3 kV is applied to overcome the potential 
difference between liquid and gas phases, then those drifted electrons 
evaporate from the liquid to the gas. 
Those electrons are further accelerated in the gas with an applied  
field of about 10$^{3}$V/cm, then electron multiplication takes
place and emits proportional lights.
The electron events can be identified with the delayed coincidence signal
from the gas phase.

Both singles from the liquid and from the gas can be measured by
photo-multiplier tubes.

In this configuration, an idea of the 'Ion Sweep' that will be discussed later
can also be possible.

\section{Backgrounds}

The backgrounds should be removed as much as possible and the remaining 
backgrounds should be well understood 
since there is no coincidence information avaiable
for the neutrino electron scattering experiments.
The backgrounds come from the impurity of the target material(Xe), the material
surrounding the detector, muon induced radioactive products and so on.
The $2\nu$ double beta decay of $^{136}Xe$ is the most serious backgrounds for 
this experiment.

\subsection{Internal backgrounds}

\subsubsection{Cosmogenic}

Fortunately Xe does not have an isotope with a lifetime longer than a couple 
of months. 
The longest isotope is $^{127}$Xe with the half-life of 36.4 days, which decay
through electron capture with the Q-value of 0.664MeV. The second longest 
isotope is $^{131}$Xe$^{m}$ with the lifetime of 11.8days. 
Therefore the cosmogenic production of the Xe isotopes is not a serious 
problem.  

\subsubsection{$^{85}Kr$, $^{42}Ar$}

Krypton and Argon may remain in the Xenon, since their boiling temperatures  
are lower than Xenon. In the nuclear reactors and from other 
sources, the radioisotopes, $^{85}$Kr, $^{42}$Ar and $^{39}$Ar are produced.
The half life of the $^{85}$Kr is 10.7 years. 
The end point energy of the beta decay is 687keV.
The current abundance, $^{85}$Kr/Kr, is 
reported to be $\sim$2$\times$10$^{-11}$. This number is also confirmed by the
direct beta decay measurement of $^{85}$Kr in liquid Kr~\cite{Kr85-II}. 
This abundant Krypton gives about 1 Bq/m$^{3}$ in air 
(Kr/Air=1.1$\times$10$^{-4}$Vol\%).
The standard Xe gas for an industrial production contains about 10ppm of Kr
which gives about 10Hz of $^{85}$Kr decays for 1 litter of liquid Xe.
Suppose we allow 1 $^{85}$Kr decay per day in 10 ton Xe detector, 
then we need to reduce the Kr in Xe down to 4$\times$10$^{-15}$g/g. 

The $^{42}$Ar ($\tau_{1/2}$=33y) concentration is evaluated to be 
$^{42}$Ar/Ar = 
7$\times$10$^{-15}$~\cite{42Ar-I,42Ar-II,42Ar-III} and gives about 
1 Bq/m$^{3}$ in air. Suppose we require that the backgrounds from this source 
is less than 1 event per day, then the Ar/Xe ratio should be less than
1.7$\times$10$^{-11}$g/g, a level easier than the $^{85}$Kr case.

Since Xe can be handled in three phases, gas, liquid and solid, there are 
variety of methods to purify Xenon.
Bubbling can be used particularly to remove Kr and Ar and may be Rn.
Helium and nitrogen are the best suited for the bubbling gases.
Distillation can also applied to separate Kr and Ar.

There is a development of the absorption column~\cite{Ab-Column}. 
They have demonstrated that
10ppm Kr contamination in Xe-gas reduced down to 1ppb 
by passing through this column. 
We expect to reduce further by applying this column repeatedly.

We should note that Xe can be circulated through the purification system and
improvements of purification systems are possible even during the experiment.

\subsubsection{$^{3}$H}
The end point energy of 18keV is sufficiently small for the solar neutrino
measurements, although $^{3}$H can be removed effectively by 
chemical processes.

\subsubsection{$^{40}$K}
$^{40}$K($\tau_{1/2}$=1.28$\times$10$^{9}y$) decay through $\beta^{-}$(89.3\%) with 
end point energy of 1.31MeV and through EC(10.7\%)with the Q-value of 1.51MeV.
The K contamination in Xe should be less then 3.9$\times$10$^{-14}$g/g if the 
$^{40}$K decay is allowed to be less than 1/day.

\subsubsection{U/Th contamination}
\label{sec-U/Th}

The radioactive decay of the U/Th chain is a large source of backgrounds 
although the solid U/Th can be removed relatively easily in the Xenon gas.
If we allow one $^{238}U$ decay per day, which produces 
8 $\alpha$'s and 6 $\beta$'s assuming in equilibrium, the corresponding 
contamination level is 9.3$\times$10$^{-17}$g/g for 10 tons of 
liquid Xenon. Similar level is required for Th-chain. 
Borexino group has achieved the U/Th contamination level of
$\sim$4$\times$10$^{-16}$g/g in their liquid scintillator.

Alpha decay can be identified by the pulse shape
discrimination or in the two phase detector as described before. 
Positive quantitative identification of U/Th contamination is possible by 
detecting Bi-Po delayed concidence, which enable us to evaluate the total 
backgrounds
coming from the U/Th contamination and to make a statistical subtraction 
of the remaining backgrounds.

In addition to that, the idea of Ion Sweeper, which is described in 
the following paragraph, may improve the situation significantly.

\subsubsection{Positive Ion Sweeper}
\label{sec-ionsweep}

Suppose the daughter nuclei of the decay products are ionized, those ions may
be sweeped away by applying the suitable electric fields in liquid Xenon 
detectors. 
Once this ion sweeper works,
all the daughter nuclides newly produced are sweeped away. 

For a first few months after the electric field turned on 
we still get $\beta$'s and $\gamma$'s from the short lived 
nuclides in the U/Th chain, but after a few months of the sweeping time, 
in the liquid-Xe, all the short lived nuclides are decayed out and only the 
long lived nuclei remain as background sources.

For the U/Th chain, the long life nuclei (\(>\)30days), $^{232}$Th, $^{238}$U, 
$^{234}$U, $^{230}$Th, $^{226}$Ra, do alpha decay. 
The alpha decay can be identified in the liquid Xenon detector by
the pulse shape discrimination.
The $^{228}$Ra (5.75yr) and $^{210}$Pb (22.3yr) are the only long life nuclei,
which
decay through $\beta$/($\gamma$).
However the Q-value is small, 45keV for $^{228}$Ra and 63keV for $^{210}$Pb.
Those backgrounds may determine the energy threshold of the experiment. 

If the ion sweep is demonstrated to work, the U/Th contamination 
would not 
be a big problem. And it is probably not necessary to achieve 
10$^{-16}$(U/Th)g/g level.

However, by a general argument, those U/Th contamination should
be as small as possible anyway, which makes the energy threshold hopefully 
lower.

Anyway the two crucial issues on the ``positive ion sweeper'' that; 
1) if the daughter nuclides are ionized and
2) if the positive ions have long enough life to drift out the chamber,
must be proved.

The ionization potential of Xe(12.13eV) is the deepest compare to the products of
U/Th-decay chain---U(6.19eV), Pa(5.89eV), Th(6.31eV), Ac(5.17eV), Ra(5.28eV),
Rn(10.75eV), At(?), Po(8.41eV), Bi(7.29eV), Pb(7.42eV) and Tl(6.11eV).
Therefore the probability for the daughter nuclides to excharge charge with Xe
is negligible during the drift.
The ion mobility in liquid Xenon has actually measured by 
\cite{IonSweep-I,IonSweep-II}.
An idea of collecting Ba ions through electric
fields was discussed in \cite{IonSweep-II} for the $\beta\beta$ decay identification.
The drift velocity of Ba ions is estimated to be 3$\times$10$^{-4}$cm$^{2}$/V/s.

We could test the this idea probably by using a small test set up.

\subsection{muongenic/spallation}

It is difficult to estimate the backgrounds produced in the liquid Xe 
by penetrating
muons at underground. The muon rate at Kamioka site for the 10 ton liquid Xe 
detector (1.8m in diameter) is $\sim$250 muons/day. 

Obviously, those muons and the prompt interactions can be removed very easily.
Short live backgrounds can be removed by so called spallation cuts, \`{a} la 
Super-Kamiokande.

Problem comes from long-lived products.
The muongenic cross section for some heavy elements like Gd, Yb, were measured 
by the accelerator experiment~\cite{Spall-Cross}. We would use their results to estimate the cross 
sections on Xe. The most cross sections for 
100 GeV muons are 9.3$\pm$1.0 mb for Gd case 
in ($^{160}$Gd$\rightarrow^{159}$Gd) and 27.7$\pm$3.9mb for Yb case 
in the reaction ($^{176}$Yb$\rightarrow^{175}$Yb).
The most cross sections come from the ($\gamma$, n) reactions and 
the ($\gamma$, 2n)
reactions, which produce (Z, A-1) and (Z, A-2) nuclides.
For Xenon case, the candidate transitions are 
$^{136}$Xe(8.87\%)$\rightarrow^{135}$Xe,
$^{134}$Xe(10.44\%)$\rightarrow^{133}$Xe,
$^{128}$Xe(1.919\%)$\rightarrow^{127}$Xe,
$^{126}$Xe(0.089\%)$\rightarrow^{125}$Xe,
$^{124}$Xe(0.096\%)$\rightarrow^{123}$Xe, and
$^{129}$Xe(26.4\%)$\rightarrow^{127}$Xe,
where, in the parenthesis, the natural abundance ratio is shown.
We may ignore of the transition of $^{124,126,128}$Xe due to the small 
abundance.
$^{127}$Xe decays through electron capture with 664keV, which could be 
identified. 
$^{135}$Xe decays for  96\% of the time with $\beta$ and $\gamma$(249keV). 
For the most of the case, the total amount of the energy released is above 
the pp-neutrinos (260keV), but it still be a problem.
$^{133}$Xe also decays through $\beta$ and $\gamma$, with 81 keV $\gamma$ 
energy, and the time difference is 6ns, which may be identified by the delayed coincidence.

As a crude guide line, suppose the muon cross section on $^{136}$Xe and 
$^{134}$Xe is $\sim$10mb,
the rate of the production of $^{135}$Xe and $^{133}$Xe is 1.9 
events/day altogether.
We have used the fact that the energy dependence of the spallation cross section, roughly proportional
to E$_{\mu}^{0.75}$~\cite{Spall-Cross,Spall-Edep,Spall-Edep-I,Spall-Edep-II}, 
give roughly factor of two increase of
the spallation cross section to E$_{\mu}\sim$285GeV for Kamioka site. 

Here we have not considered other channels to produce neighbouring element 
like I, Cs and so on. Which may have a contribution of the similar order.
The background situation of muongenics is probably marginal.

If we go about 100m deeper than the Super-Kamiokande level, 
then the muon rate reduces by 1order of magnitude therefore the problem would 
disappear. Since the size of the cavity required for the experiment is 
small, therefore there is no trouble to find and go deeper places.

Further reduction of the spallation backgrounds would also be achieved:
\begin{itemize}

\item 
If the ion sweep, which is described in ~\ref{sec-ionsweep} 
as a method to sweep the decay products of U/Th-chains, 
is demonstrated to work, then those spallation products are also sweeped 
out, since the spallation products are supposed to be ionized.

\item 
If we adopt the isotope separation of $^{136}$Xe(removing), 
to be discribed later the most spallation problem ($^{136}$Xe $\rightarrow^{135}$Xe) disappears.

\end{itemize}

Anyway it is important and necessary to measure the spallation cross sections 
by muon beam produced by high energy accelerators.
 
\subsection{External backgrounds}

The $\gamma$-rays and neutrinos from the materials surrounding the 
liquid Xenon---container, PMTs, cables, structures and rocks---would be 
backgrounds. 
The shell structure, which provide a graded shielding against those 
radio-activites, would reduce the backgrounds.

The inner structures would require more purity than those situated outer side
of the detector.

The high density and high Z of Xenon
would provide very good self-shield against those background. The 30cm 
self-shield is equivalent to the 4m of water shield for the 
electro-magnetic component. The detailed study and simulations for background 
should be done.

\subsection{Backgrounds from double beta decay}

\begin{figure}
\begin{center}
\psbox[height=12.0cm]{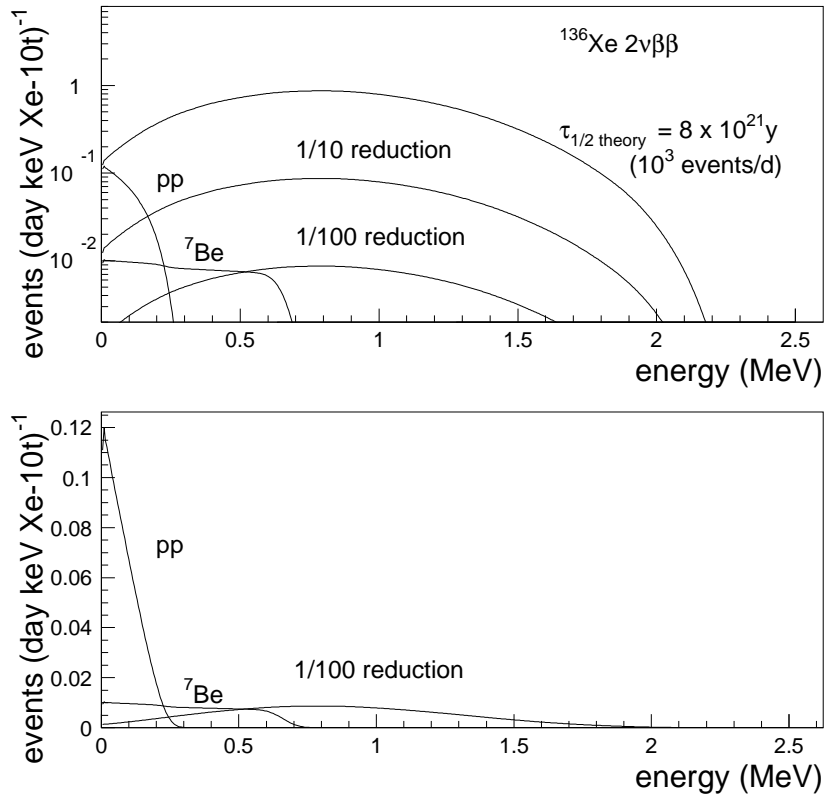}
\caption{Background from 2$\nu$-double $\beta$ decay of $^{136}$Xe.}
\label{fig:Double-b-BG}
\end{center}
\end{figure}

Xenon contains two double beta-decay and two double beta-plus decay
isotopes.
Among them the $^{136}$Xe 2$\nu$-double beta decay is a potential problem,
since the Q-value(2.48MeV) is highest and the natural abundance(8.9\%) is high.
The expected event rate assuming the 8$\times$10$^{21}$yr for 2$\nu$ double 
beta decay is about 1,000 events per day as shown in Fig.~\ref{fig:Double-b-BG}. 
The energy spectrum extends up to 
2.48 MeV.  
We expect 10 to 20pp- solar neutrino interactions below 300 keV where 
about 10\% of the double beta signals, about 100 events, overlaps.
If the lifetime of the double beta decay is longer(say 100 times of 8$\times$
10$^{21}$ yr), then the solar neutrino 
signal may extracted and the double beta decay would not be a problem.
It is necessary to determine the lifetime of the 2$\nu$ $\beta\beta$-decay.
The current experimental lower limit is 0.5$\times$10$^{21}$yr.
Of course the 10 ton Xenon (0.87 tons of $^{136}$Xe) has a potential sensitivity 
for 0$\nu$ double beta decay.
If the lifetime of the double beta decay is shorter, then the isotope 
separation of $^{136}$Xe is inevitable.
The isotope separation will be essential and improve the situation 
significantly, which will be discussed in the next section~\ref{sec-isotope}.

\section{Isotope separation and detection of 
solar neutrinos, double beta decay and dark matter.}
\label{sec-isotope}

The isotope separation (enrichment or depletion) would be essential for this 
experiment, since it is very difficult to anticipate the lifetime of the 2$\nu$
double beta decay to be 100 times longer than the current theoretical estimate 
of 2$\sim$8$\times$10$^{21}$yr.

But, once we accept the necessity of the isotope separation, then
we will obtain other possible and interesting applications than the solar 
neutrino detection:
One should note that it will open up an new possiblity of a positive 
identification of the dark matter detection. 
The isotope distribution of Xenon is, $^{124}$Xe(0.096\%), $^{126}$Xe(0.089\%),
$^{128}$Xe(1.919\%), $^{129}$Xe(26.4\%), $^{130}$Xe(4.07\%), 
$^{131}$Xe(21.18\%), $^{132}$Xe(26.89\%), $^{134}$Xe(10.44\%),
$^{136}$Xe(8.87\%). 
If we are able to split between 131 and 132, we can 
separate Xe into two sub-sets which consist of mostly odd nucleus 
(129(26.4\%) and 131(21.2\%)) 
and mostly even nucleus (132(26.9\%), 134(10.44\%) and 136(8.87\%)). 
Of course it is not possible to make a clear separation, but anyway we are 
able to measure the dark matter interaction separately with the mostly-odd
nuclei and with the mosly-even nuclei to extract the information of spin 
independent and spin dependent interactions. Futhermore, for example, by
exchanging the detector containers of two sub-sets, we can set the background
environment to be common for the two sub samples which is very important.

Therfore the experiment can be configured dynamical way:  
the solar neutrinos detection can be done by odd enriched sample and 
the double beta decay experiment can be done
by even enriched sample and dark matter can be measured in both samples.
The isotope separation really gives a new experimental approach and 
opportunities for a variety of physics.

In addition to the increased physics opportunities, the purity of liquid 
Xenon can be significantly improved by the isotopr separation.

\section{Conclusions}

The 10ton liquid Xenon solar neutrino detector by which we expect to detect
$\sim$14pp and $\sim$6 $^{7}$Be neutrinos per day can be built by known 
technologies.
The impurity level of the liquid Xe should be reduced down to 10$^{-16}$g/g, 
10$^{-15}$g/g and 10$^{-11}$g/g for U/Th, $^{85}$Kr and $^{42}$Ar, 
respectively. Alpha decay may further be identified and be separated 
from the electron events. 
The required purity level can be accomplished by bubbling, distillation, 
filtering, and so on.
The detector is shielded by the outer 30cm layer of Xe against the incoming 
backgrounds.
Spallation products are estimated to be about 2 events per day.
The ion sweeper, if possible, improves the situation.

Most serious background comes from the 2$\nu$ double beta decay of $^{136}$Xe. 
The lifetime of the 2$\nu$ double beta decay should be determined before the
final design of the experiment, which infers 
the required level of the isotope separation.

The isotope separation is inevitable if the 2$\nu$ lifetime is shorter than 
100$\times$ 8$\times$ 10$^{21}$yr.

The isotope separation, however, improves the situation in many aspects, 
not only for the backgrounds from the double beta decay, but also for the 
internal background.

Xe could be divided into odd enriched and even enriched 
samples by the isotope separation. 
Two detectors can be prepared for the two samples,
and the odd enriched sample, which does not 
contain the large amount of $^{136}$Xe can be used for the solar neutirno 
detection and the even enriched sample can be used for the double beta decay 
study.
An occasional exchange of the containers would provides the way to evaluate 
the background in each sub sample and provide a new way to identify the dark 
matter signal by the evaluation of the spin independent and dependent 
interations through the separately measured interactions with odd and even 
nuclei.
  
\vspace{0.5cm}
We acknowledge Stefan Schoenert for many stimulated discussions.
This work was partly supported by the Grant-in-aid on Scientific 
Research by the Japanese Ministry of Education, Science and Culture.

\newpage
\begin{table}[htb]
\begin{center}
\begin{tabular}{r|r|r|r|r|r}  
                         & 
 He	& Ne	& Ar	& Kr	& Xe	\\ 
atomic number		&
 2	& 10	& 18	& 36	& 54	\\
atomic mass	&
4.00260	&20.1797&39.948	&83.80	& 131.29\\
Ionization Potential(eV)	&
 24.587	& 21.564	& 15.759	& 13.999	& 12.13	\\
operating temperature(boiling point)(k)              &
 4.215	& 27.096	& 87.28         & 119.8 	& 165.03 \\
wave length of scintillation light                       & 
 73nm	& 80nm	& 128nm	& 147nm	& 174nm	\\
number of photons(/MeV)            &
 	& 	& 40,000 & 	& 43,000 \\
Long life Isotopes(\(>\)a months)                &
--- 	& ---	 & $^{39}$Ar,$^{42}$Ar & $^{85}$Kr & --- 	 \\
density in liquid phase(g/cm$^3$)  &
 0.125	& 1.20	& 1.40	& 2.6	& 3.06	\\
radiation length(cm)	&
756	& 24	& 14	&	& 2.4	\\
Volume fraction in air (in \%)		&
0.0005	& 0.0018 & 0.9325 & 0.0001 & 0.0000009 \\
\end{tabular}
\caption{Comparison between noble gas scintillators.Source Particle data book
and others}
\label{tbl:comp}
\end{center}
\end{table}


\newpage

\end{document}